\documentclass{emulateapj}

\usepackage{psfig}
\usepackage{amsmath}
\usepackage{amssymb}
\usepackage{graphicx}

\newcommand{\MCcode}{\texttt{SEDONA}}
\newcommand{\kms}{\ensuremath{\mathrm{km~s}^{-1}}}

\newcommand{\Mni}{\ensuremath{M_{\mathrm{Ni}}}}
\newcommand{\Nifs}{\ensuremath{^{56}\mathrm{Ni}}}
\newcommand{\Cofs}{\ensuremath{^{56}\mathrm{Co}}}
\newcommand{\Fefs}{\ensuremath{^{56}\mathrm{Fe}}}

\newcommand{\dmfb}{\ensuremath{\Delta M_{15}(B)}}
\newcommand{\dmfv}{\ensuremath{\Delta M_{15}(V)}}
\newcommand{\dmfbol}{\ensuremath{\Delta M_{15}(\mathrm{bol})}}
\newcommand{\dmfbolt}{\ensuremath{\Delta M_{30}(\mathrm{bol})}}
\newcommand{\dmfbt}{\ensuremath{\Delta M_{30}(B)}}
\newcommand{\dmfubv}{\ensuremath{\Delta M_{15}(UBVRI)}}

\newcommand{\Msun}{{\ensuremath{\mathrm{M}_{\odot}}}}

\newcommand{\Mb}{\ensuremath{M_B}}
\newcommand{\Mv}{\ensuremath{M_V}}
\newcommand{\Mbol}{\ensuremath{M_{\rm bol}}}
\newcommand{\bmvmax}{\ensuremath{(B-V)_{\rm max}}}
\newcommand{\bmvf}{\ensuremath{(B-V)_{15}}}
\newcommand{\tbol}{\ensuremath{t_{\rm bol}}}
\newcommand{\tb}{\ensuremath{t_{\rm B}}}
\newcommand{\td}{\ensuremath{t_{\rm d}}}
\newcommand{\Bband}{\ensuremath{B}-band}
\newcommand{\bmv}{\ensuremath{B-V}}
\newcommand{\bmr}{\ensuremath{B-R}}

\shortauthors{Kasen \& Woosley}
\shorttitle{Type~Ia Supernova Width Luminosity Relation}

\submitted{}

\begin{document}

\title{On the Origin of the Type~Ia Supernova Width-Luminosity Relation}

\author{Daniel Kasen\altaffilmark{1,2}\email{kasen@pha.jhu.edu} and S.E. Woosley\altaffilmark{3}}
\altaffiltext{1}{Allan C. Davis Fellow, Department of Physics and Astronomy, Johns Hopkins University, Baltimore, MD 21218}
\altaffiltext{2}{Space Telescope Science Institute, Baltimore, MD 21218}
\altaffiltext{3}{University of California, Santa Cruz}

\begin{abstract}
Brighter Type~Ia supernovae (SNe~Ia) have broader, more slowly
declining \Bband\ light curves than dimmer SNe~Ia.  We study the
physical origin of this width-luminosity relation (WLR) using detailed
radiative transfer calculations of Chandrasekhar mass SN~Ia models. We
find that the luminosity dependence of the diffusion time (emphasized
in previous studies) is in fact of secondary relevance in
understanding the model WLR.  Instead, the essential physics involves
the luminosity dependence of the
\emph{spectroscopic/color} evolution of SNe~Ia.  Following
maximum-light, the SN colors are increasingly affected by the
development of numerous Fe~II/Co~II lines which blanket the \Bband\
and, at the same time, increase the emissivity at longer wavelengths.
Because dimmer SNe~Ia are generally cooler, they experience an earlier
onset of Fe~III to Fe~II recombination in the iron-group rich layers
of ejecta, resulting in a more rapid evolution of the SN colors to the
red.  The faster \Bband\ decline rate of dimmer SNe~Ia thus reflects
their faster ionization evolution.
\end{abstract}


\section{Introduction}

Although normal Type~Ia supernovae (SNe~Ia) are generally considered a
homogeneous class, they nevertheless show substantial ($\sim 1$~mag)
variations in luminosity at peak.  The use of SNe~Ia for cosmology
measurements thus relies on empirical calibration techniques.  Most
common among these is the width-luminosity relation
\citep{Phillips_1999}.  Bright SNe~Ia generally 
have broad \Bband\ light curves (LCs) which decline slowly after
peak. Dimmer SNe~Ia have narrower, more quickly declining LCs.

Understanding the physical origin of the width-luminosity relation
(WLR) is a primary goal of the theory and modeling of SNe~Ia.
Unfortunately, a full theoretical description of the SN~Ia explosion
mechanism is still lacking.  Normal SNe~Ia are widely believed to be
the thermonuclear disruption of carbon-oxygen white dwarfs near the
Chandrasekhar limit, however a number of uncertainties remain
regarding the structure of the progenitor, the precise ignition
conditions, and the physics of the turbulent nuclear combustion that
unbinds the star.

Fortunately, these uncertainties need not be fully resolved in order
to study the WLR. In SN~Ia explosions, hydrodynamic and nuclear
burning processes last only $\sim 1$~minute.  The subsequent
luminosity is powered entirely by the decay of radioactive elements
synthesized in the explosion, in particular
\Nifs\ in the decay chain $\Nifs \rightarrow\ \Cofs \rightarrow\
\Fefs$.  The LCs of SNe~Ia are therefore fully determined by the
composition and density structure of the material ejected in the
explosion.  Naturally, the mass of \Nifs\ produced (\Mni) is the
primary determinate of the peak brightness of the event.  Observations
indicate that for normal objects, \Mni\ spans the range
$0.4-0.9$~\Msun, with a typical value $\Mni \approx 0.6~\Msun$.  

The challenge thus falls to radiative transfer theory to explain why
SNe~Ia with larger \Mni\ also have broader \Bband\ LCs.  Most previous
transfer studies have emphasized diffusion arguments of one sort or
another \citep{Hoeflich_WLR, Hoeflich_99by, Pinto-Eastman_WLR,
Mazzali_WLR}.  In SNe~Ia, the ejecta remain optically thick for the
first several months after explosion.  The width of the bolometric LC
is related to the time scale for photons to escape the ejecta by
diffusion.  In principle, the WLR could be explained if brighter
SNe~Ia have higher effective opacities and hence a longer diffusion
time.  Different authors have invoked different physical arguments to
motivate this sort of opacity dependence (see
\S\ref{Sec:Others}).

In this paper, we stress that the WLR does not in fact hinge on the
bolometric diffusion time, but is instead principally a
\emph{broadband} phenomenon.   In particular, the \Bband\ LC depends 
sensitively on the rate at which the SN colors evolve progressively
redward following maximum light.  Dimmer SNe~Ia exhibit a faster color
evolution than brighter SNe~Ia; this turns out to be a primary reason
for their relatively faster
\Bband\ decline rates.  A relevant theoretical explanation of the 
WLR should therefore focus on the effect \Mni\ has on the
\emph{spectroscopic/color evolution} of SNe~Ia, rather than on the
overall diffusion time-scale.

In what follows, we use realistic time-dependent multi-group radiative
transfer calculations to explain the physical origin of the WLR in a
set of Chandrasekhar-mass SN~Ia models. We demonstrate that the faster
color evolution of dimmer models reflects their faster
\emph{ionization evolution}.  Following 
maximum-light, the SN colors are increasingly affected by the
development of numerous Fe~II/Co~II lines which blanket the
\Bband\ and, at the same time,
increase the emissivity at longer wavelengths.  Because dimmer SNe~Ia
are generally cooler, they experience an earlier onset of Fe~III to
Fe~II recombination in the iron-group rich layers of ejecta, resulting
in a more rapid evolution of the SN colors to the red.  This explains
their faster \Bband\ decline rate.

\section{Previous Theoretical Studies}
\label{Sec:Others}

Previous radiative transfer studies of SNe~Ia have occasionally
reproduced the basic trend of the WLR among sets of models with
varying \Mni.  In these studies, the model relation is typically
explained by claiming a strong luminosity dependence of the diffusion
time.  To the extent that the transfer calculations include the
relevant multi-wavelength processes, color evolution effects are also
captured in the simulations as a matter of course.  Sometimes their
relevance to the WLR is explicitly recognized and quantified
\citep[e.g.,][]{Mazzali_WLR}.

Using simple dimensional arguments, one can show that the photon
diffusion time (and hence bolometric LC width) scales with mean
opacity as $t_d \propto \kappa^{1/2}$.  In SNe~Ia, the dominant
opacities are due to electron scattering and bound-bound line
transitions, the latter of which is enhanced by Doppler broadening in
the differentially expanding ejecta.  Iron group species (iron,
cobalt, and nickel) have the largest line opacity, due to their
complex atomic structure.  The iron group line opacities increases
sharply to the blue, due to the larger number of lines at shorter
wavelengths.

At least three different (though related) arguments have been given as
to why $\kappa$ (and hence the diffusion time) should increase with
\Mni, each emphasizing a different physical mechanism.

\citet[][see also \cite{Khokhlov_93,Hoeflich_WLR}]{Hoeflich_99by} 
emphasize the temperature dependence of the diffusion time.  Because
SNe~Ia with larger \Mni\ have higher temperatures, the bulk of the
radiation energy is concentrated at shorter wavelengths.  Since the
line opacity increases sharply to the blue, the effective mean opacity
(and hence the diffusion time) increases with \Mni.

\citet[][see also \cite{Pinto-Eastman_I, Pinto-Eastman_II}]{Pinto-Eastman_WLR} 
emphasize the effect of the ionization state on the diffusion time.
They argue that \td\ is largely determined by the rate at which
blue/ultraviolet photons fluoresce to longer wavelengths via
interaction with iron group elements, thereby enhancing photon escape.
They suggest that fluorescence is more efficient in singly, as opposed
to doubly ionized species.  Because SNe~Ia with lower \Mni\ are cooler
and less ionized, the diffusion time is shorter.

\cite{Mazzali_WLR} emphasize the effect of the ejecta composition on
the diffusion time.  SNe~Ia with large \Mni\ have a greater abundance
of iron group elements, and thus presumably larger opacities.

In fact, all three of the above mentioned diffusion effects must be
operative at some level in the LCs of SNe~Ia.  The magnitude of the
effects (in sum) can be quantified by direct examination of the model
bolometric LCs.  If diffusion time arguments are indeed of prime
relevance in explaining the WLR, it follows that the bolometric LCs of
brighter models must be significantly broader than those of dimmer
models.  However, in the previous studies this bolometric behavior is
not explicitly demonstrated using detailed multi-group transfer
calculations and realistic opacities.

In the following, we present a set of model calculations in which the
shape of the bolometric LCs in fact shows very little direct
dependence on \Mni.  Variations in the diffusion time are therefore of
secondary relevance in addressing the \Bband\ WLR.  Instead, the model
WLR primarily reflects the faster spectroscopic/color evolution of
SNe~Ia with lower \Mni.  It is therefore of fundamental importance
(and a primary goal of this paper) to explain in detail the physics
controlling the rate of color evolution in SNe~Ia.

Because the flux mean opacity itself depends on the spectral energy
distribution of the radiation field, the physics controlling the color
evolution of SNe~Ia is of course deeply intertwined with that
determining the diffusion time.  In the end, however, the relative
importance of each can be quantified by comparing the strength of the
WLR in the broadband versus the bolometric LCs of SNe~Ia.  We suggest
below that such an analysis may be very useful in furthering our
understanding of SNe~Ia.

\begin{deluxetable*}{llllllllllllll} 
\tablecaption{Observable properties of the models}
\tablehead{
\Mni & $t_{bol}$ & $t_B$ & $M_{\rm bol}$ & $M_B$ & \Mv\ & \dmfbol\ & \dmfb\ & \dmfv\ & \dmfubv & \bmvmax & \bmvf\  }
\startdata
0.70   &  15.56   &  19.34   &  -19.46   &  -19.34   &  -19.35   &  0.81   &  1.09   &  0.72   &  0.83 &    -0.017 &    0.568  \\
0.63   &  15.55   &  19.00   &  -19.36   &  -19.25   &  -19.27   &  0.82   &  1.17   &  0.77   &  0.84 &    -0.007 &    0.621  \\
0.56   &  15.54   &  18.70   &  -19.24   &  -19.14   &  -19.18   &  0.83   &  1.28   &  0.82   &  0.88 &    0.011 &    0.693  \\
0.49   &  15.56   &  18.36   &  -19.11   &  -19.02   &  -19.08   &  0.84   &  1.38   &  0.86   &  0.91 &    0.034 &    0.750  \\
0.42   &  15.52   &  17.85   &  -18.96   &  -18.87   &  -18.97   &  0.84   &  1.45   &  0.92   &  0.94 &    0.057 &    0.802  \\
0.35   &  15.51   &  17.22   &  -18.78   &  -18.71   &  -18.83   &  0.84   &  1.53   &  0.98   &  0.96 &    0.089 &    0.851  \\
\enddata
\end{deluxetable*}

\section{Models}

In a companion paper \citep[][hereafter Paper~I]{Woosley_WLR}, we
perform an extensive parameter study of SN~Ia LCs using 1-dimensional
(1D) hydrodynamical models of the explosion.  Here we focus on a
restricted set of those models in order to demonstrate the basic
transfer physics underlying the WLR.  Each model is a derivative of
the spherical Chandrasekhar mass explosion model, M070103, which has a
four zone stratified compositional structure.  From the center out,
these zones consist of: 0.1~\Msun\ of stable iron group elements;
0.7~\Msun\ of \Nifs; 0.3~\Msun\ of intermediate mass elements (IMEs;
silicon,sulfur, calcium, argon); and 0.3~\Msun\ of unburned
carbon/oxygen.  Mild mixing is applied at the zone interfaces.  The
synthetic LCs and spectral time-series of model M070103 are in good
agreement with observed normal SNe~Ia (Paper~I).

A simple prescription is used to vary the \Nifs\ abundance in model
M070103, allowing us to isolate the effect of \Mni\ on the LCs.  New
models are constructed by reducing the abundance of \Nifs\ everywhere
by a fractional amount (viz., $10,20,30,40,$ and $50$\%) while
increasing the abundance of IMEs by a corresponding amount.  This
amounts to a variation of \Mni\ from 0.35 to 0.7~\Msun\ while
maintaining a fixed velocity distribution of \Nifs.  Because the
energy released in burning to IMEs is comparable to burning to \Nifs,
the dynamics of the explosion will not be greatly affected by the
compositional change.  The ejecta structures bear resemblance to some
1D delayed-detonation or pulsed-detonation models
\citep{Hoeflich_DD}.

We compute synthetic LCs of the models using the multi-dimensional
time-dependent Monte Carlo radiative transfer code \MCcode\
\citep{Kasen_MC}.  Given a homologously expanding SN ejecta structure,
\MCcode\ calculates the emergent spectral time series at high
wavelength resolution.  Broadband LCs are then constructed by
convolving the spectra with the appropriate filter transmission
functions. The code includes a detailed gamma-ray transfer procedure
to determine the rate of radioactive energy deposition, and a
radiative equilibrium solution of the temperature structure.  No
ad-hoc inner boundary condition is employed.  Ionization and
excitation are computed assuming local thermodynamic equilibrium, and
bound-bound line transitions are treated using the expansion opacity
formalism and an approximate two-level atom approach to wavelength
redistribution, assuming a constant redistribution probability
$\epsilon = 0.8$ for all lines. We have used
\MCcode\ to recalculate the broadband LCs of the scaled models used to
study the WLR in
\cite{Pinto-Eastman_WLR} and found excellent agreement with the
results presented there.  See \cite{Kasen_MC} for a detailed code
description and verification.

\section{Origin of the Width-Luminosity Relation}

\begin{figure}
\begin{center}
\includegraphics[width=8.5cm]{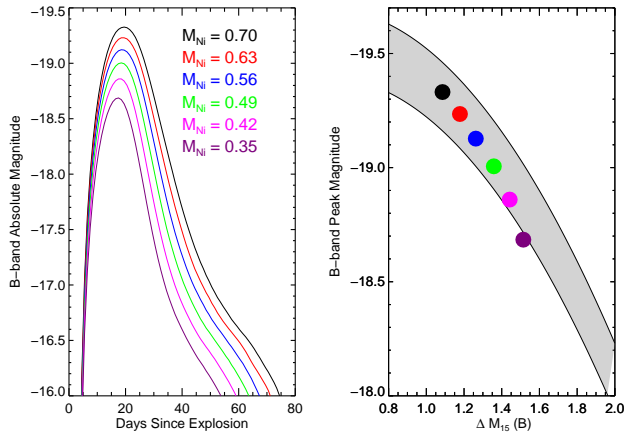}
\caption{\Bband\ light curves (left panel) and width-luminosity relation
  (right panel) of the models discussed in the text. The shaded region
  in the right panel is the empirical WLR of \cite{Phillips_1999} with
  a magnitude dispersion of 0.15~mag and absolute calibration $\Mb =
  -19.3$~mag at $\dmfb = 1.1$~mag.
\label{Fig:wlr}}
\end{center}
\end{figure}

The WLR is often quantified as a relation between peak \Bband\
magnitude, \Mb, and the drop in \Bband\ magnitude 15 days after peak,
\dmfb.  Table~1 lists the relevant observable parameters of the
synthetic LC models discussed in the previous section.  In the models,
the variation of \Mni\ from 0.35 to 0.70~\Msun\ results in a variation
in
\Mb\ from -18.71 to -19.34~mag, and in \dmfb\ from 1.53 to 1.09~mag.
As seen in Figure~1, the models obey a clear WLR, with a slope that
agrees reasonably (though not exactly) with the observed one of
\cite{Phillips_1999}.

\begin{figure}[t]
\begin{center}
\includegraphics[width=8.5cm]{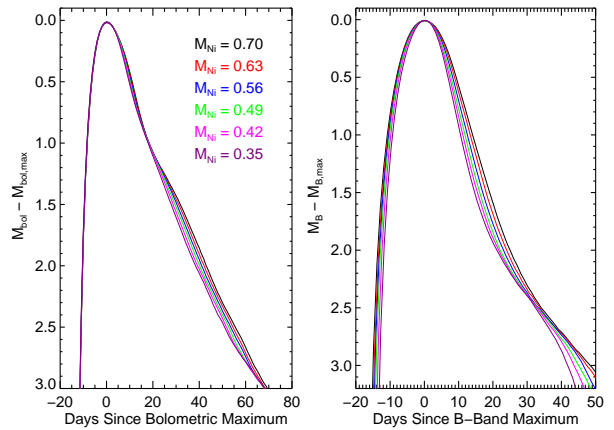}
\caption{\emph{Left:} 
Bolometric LCs of the models, normalized at peak and shown relative to
time of bolometric maximum.  The LC shapes are all very similar over
the first three weeks after maximum light.  Moderate differences of
order 0.15~mag appear around 30 days after maximum, when the brighter
models reach a small secondary maximum.
\emph{Right:} \Bband\ of the models, normalized at peak and shown
relative to time of $B$-maximum.  The LC decline rate just after maximum
depends strongly on
\Mni.
\label{Fig:bol}}
\end{center}
\end{figure}

The model WLR obtained in Figure~\ref{Fig:wlr} is not attributable to
a strong systematic dependence of the diffusion time on \Mni, as
emphasized in previous studies.  On the contrary, the bolometric rise
time to maximum and the bolometric decline rate are similar among all
models: $\tbol = 15.51-15.56$~days, $\dmfbol = 0.81-0.84$~mag.  As
seen in Figure~\ref{Fig:bol}, differences in the bolometric LCs are
maximal (though still modest) when measured 30 days after $B$-maximum:
$\dmfbolt = 1.43-1.61$~mag.  Thus, in this set of models, the
luminosity dependence of the diffusion time is in fact of secondary
relevance in understanding the WLR.

Instead, we identify the crucial physical effect to be that of \Mni\
on the spectroscopic/color evolution of the models.  As seen in
Figure~\ref{Fig:colors}, all models have very similar colors at
\Bband\ maximum, with \bmv\ and \bmr\ within $0$ and $0.15$~mag.
Thereafter, the color curves diverge, with the dimmer models evolving
redward more rapidly.  By 15 days after $B$-maximum, differences in
\bmv\ and \bmr\ color have grown to nearly 0.3 and 0.6~mag
respectively.  It is this faster color evolution of the dimmer models
that accounts for their larger \dmfb.

The origin of this color evolution behavior can be traced directly to
the impact of iron-group lines on the post-maximum spectra. Soon after
maximum light, the \Bband\ is increasingly affected by the development
of Fe~II/Co~II lines (Figure~\ref{Fig:evolution}).  Especially
prominent are the strong absorption blends at wavelengths
4000-4600~\AA\ and 4800-5300~\AA, however a pseudo-continuum of
numerous weaker lines also contribute to the overall line blanketing
in the blue.  At the same time, a large number of blended Fe~II/Co~II
lines emerge in the far red and near-infrared, enhancing the
pseudo-continuum emissivity at these wavelengths
\citep{Kasen_NIR}.  The net effect of the onset of the
Fe~II/Co~II lines is thus to redistribute flux absorbed in the blue
part of the spectrum to longer wavelengths.  Naturally, the rate at
which these lines develop plays a critical role in determining the
color evolution (and hence
\Bband\ decline rate) of SNe~Ia.

In general, Fe~II/Co~II lines become prominent once the layers of
ejecta rich in \Nifs\ (since decayed to iron/cobalt) begin to
recombine from doubly to singly ionized, a process which occurs quite
suddenly at temperature $T \approx 7000$~K.  For the epochs at and
before \Bband\ maximum, the iron/cobalt-rich layers of ejecta ($v \la
9000$~\kms) have $T > 7000$~K and Fe~II/Co~II lines are weak or absent
in all model spectra (see Figure~\ref{Fig:spectra}; top panel).  At
these times, the optical continuum is pseudo-blackbody and the
spectral and color differences among the models are minor, even when
\Mni\ is varied by a factor of two.  The brighter/broader SNe~Ia have
a bluer
\bmvmax\ color, but by a relatively small amount, as noted in
observations \citep{Phillips_1999}.

In the weeks following maximum light, however, the layers of iron-rich
ejecta progressively cool to $T \la 7000$~K and, as these layers
recombine, the Fe~II/Co~II lines become steadily stronger.  The exact
rate at which the lines develop depends sensitively on the overall
ejecta temperature scale.  Models with lower \Mni\ have generally
lower temperatures and hence an earlier onset of Fe~III to Fe~II
recombination.  As seen in Figure~\ref{Fig:spectra} (bottom panel) by
15 days after maximum the line blanketing is much stronger (and the
colors much redder) in the lower \Mni\ model.  The larger
\dmfb\ of the dimmer model  thus reflects the  
more rapid development of the Fe~II/Co~II lines.

To be clear, the WLR is not merely a consequence of the generally
redder colors of dimmer SNe~Ia, but rather of the faster rate at which
the colors \emph{evolve} in these objects.  In considering this
spectroscopic effect, it is useful to keep in mind that the
temperature differences among normal SNe~Ia are expected to be rather
small.  The ejecta temperature (as determined by the balance of
radiative heating and cooling) scales with luminosity as $T
\propto L^{1/4}$ and hence $T \propto
\Mni^{1/4}$. A factor of two difference in \Mni\ therefore amounts to
only a $\sim 20\%$ difference in temperature.  If the SN spectrum is
reasonably characterized by a blackbody, the corresponding variations
in \bmv\ or \bmr\ color would be relatively small ($\la 0.15$~mag).
Indeed, this is essentially why the pre-maximum and maximum light
colors of normal SNe~Ia show only a mild dependence on \Mni.

In the post-maximum epochs, however, the SN~Ia spectrum increasingly
deviates from blackbody due to the burgeoning strength of the
Fe~II/Co~II lines.  The strength of these lines depends very
sensitively on the ionization state.  The ionization in turn obeys an
extremely non-linear dependence on temperature, undergoing sudden
recombination at $T \approx 7000$~K.  Thus, in the post-maximum
epochs, small differences in temperature translate to large
differences in \bmv\ and \bmr\ color.  The onset of this non-linear
behavior is why the decline rate \dmfb\ offers a good diagnostic of
the SN temperature scale, and hence luminosity.  Other approaches to
calibrating SNe~Ia, such as the CMAGIC method \citep{Wang_Cmagic} and
the $C_{12}$ correlation
\citep{Wang_C12} are likewise explained by the same strong dependence
of the SN colors on the ionization state during the weeks following
maximum light.

At around 30 days after $B$-maximum, the iron group layers of ejecta
finish completely recombining from doubly to singly ionized.  From
this time on, differences in the ionization state among the models
cease to be of great significance.  The color curves reconverge and
the models enter a phase of nearly identical color evolution (see
Figure~\ref{Fig:colors}).  The same general behavior is noted in the
observations \citep{Lira_1996, Phillips_1999}.  For this reason, the
decline rate measured over longer time periods (e.g., \dmfbt) provides
a less sensitive measure of the model peak magnitude.

\section{Relevance To Observations}

\begin{figure}[t]
\begin{center}
\includegraphics[width=8.5cm]{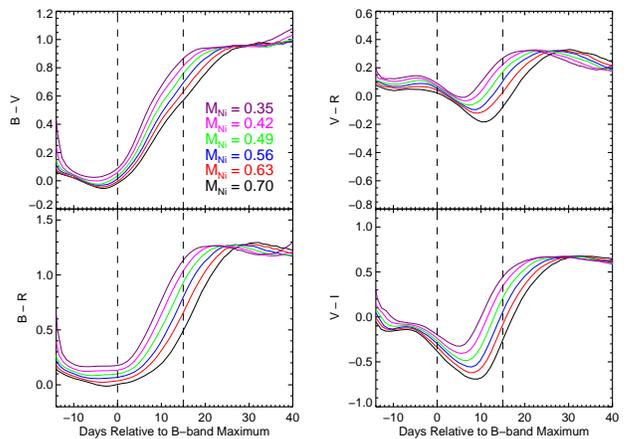}
\caption{Color evolution of the models discussed in the text.  In all panels, 
the dimmer models are the topmost lines. All models have similar
colors near maximum light; thereafter the dimmer ones (lower \Mni)
become redder more rapidly.  The differing rates of color evolution
account for differences in \dmfb.
\label{Fig:colors}}
\end{center}
\end{figure}

\begin{figure}
\begin{center}
\includegraphics[width=8.5cm]{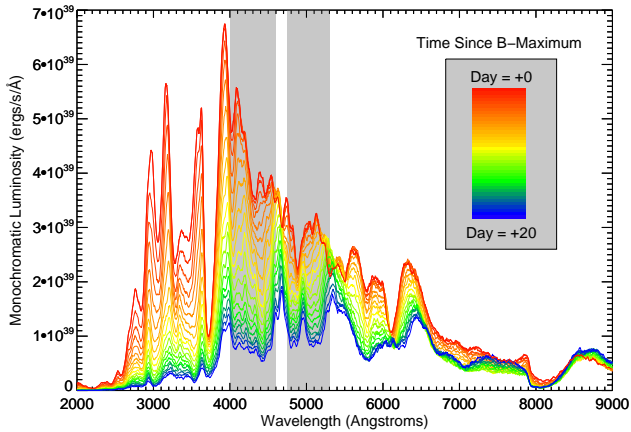}
\caption{Spectral evolution from \Bband\ 
maximum to twenty days later for the $\Mni = 0.7~\Msun$ model.  The
figure illustrates how the drop in $B$-band magnitude depends
sensitively on color changes due, in large part, to the development of
spectral line blanketing from iron group elements. The shaded bands in
the plot highlight two prominent spectral absorption features due to
Fe~II/Co~II line blends.
\label{Fig:evolution}}
\end{center}
\end{figure}

\begin{figure}
\begin{center}
\includegraphics[width=8.5cm]{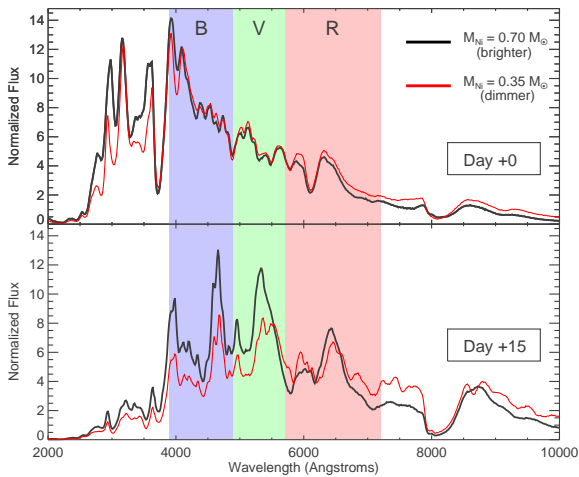}
\caption{Demonstration of the crucial effect of Fe~II/Co~II
line blanketing on \dmfb.  The figure compares synthetic
spectra of the \Mni = 0.35 and \Mni = 0.70~\Msun\ models.  All spectra
have been normalized by dividing by the bolometric luminosity at the
time. The blue, green, and red color shadings show the regions of the
B,V and R bands respectively.  Near maximum light (top panel) both
models have similar, pseudo-blackbody coninuua.  Fifteen days later
(bottom panel) broad absorption features from Fe~II/Co~II lines
dominate the \Bband.  Because the cooler \Mni = 0.35~\Msun\ model is
less ionized, the post-maximum line blanketing develops more quickly,
leading to the faster \Bband\ decline rate.
\label{Fig:spectra}}
\end{center}
\end{figure}

The observed color evolution of SNe~Ia bears a striking similarity to
the model behavior of Figure~\ref{Fig:colors}
\citep[e.g.,][Figure~1]{Wang_C12}.  Normal SNe~Ia display fairly
uniform \bmv\ colors at maximum light; thereafter the dimmer objects
become redder more quickly.  The observed differences in color at day
$+15$ are comparable to the differences in \dmfb.

\cite{Gaston_EW} 
quantifies the spectral evolution of SNe~Ia directly using equivalent
width measures.  His approach nicely illustrates the sudden growth of
Fe~II/Co~II features (there labeled ``Mg~II'') in the $B$-band
beginning just after maximum light.  The rate at which these features
develop is shown to correlate tightly with \dmfb.  Such analyses
clearly point to the spectroscopic nature of the WLR.

The construction of bolometric LCs from $uvoir$ observations of SNe~Ia
can be used to further explore this issue.  If the luminosity
dependence of the diffusion time is in fact central to explaining the
WLR, one expects to observe a strong width luminosity relation in the
\emph{bolometric} LCs of SNe~Ia.  It is currently unclear from the limited
observational sample whether such a relation exists at all.
\cite{Phillips_99ac}, for example, construct bolometric LCs 
from $UBVRI$- and (when available) $JHK$-band observations of 15
SNe~Ia.  The bolometric decline rates \dmfbol\ of the sample show some
degree of scatter, but no clear correlation with the peak bolometric
magnitude $M_{\rm bol}$.

The bolometric LC constructions of
\cite{Contardo_bol}, which use primarily $BVRI$-band observations, do show a
noisy correlation between
\dmfbol\ and \Mbol\ \citep[see also][]{Stritz}.  
This correlation, however, may in large part reflect the lack of
ultraviolet and near-infrared (NIR) data used in the bolometric
constructions.  At day 15 after $B$-maximum, for example, 20\% or more
of the SN~Ia flux may occur at NIR wavelengths
\citep{Suntzeff_bol}.  As the NIR contribution at day~15
is relatively greater in the dimmer SNe~Ia, $UBVRI$ constructions may
systematically overestimate the value of \dmfbol\ for dimmer objects
by as much as $0.1-0.2$~mag.  This behavior is seen in our models
(Table~1) where one notes that
\dmfubv\ exhibits a much stronger dependence on
\Mni\ than does the true bolometric decline rate 
\dmfbol.  In fact, the slope of our
$UBVRI$ model WLR agrees fairly well with the empirical one of
\cite{Contardo_bol}, suggesting that the 
observed correlation reflects, in large part, the role of color
evolution.

Clearly the construction of true bolometric LCs for a sample of SNe~Ia
using UV through $K$-band observations would be of great value in
constraining the radiative transfer in SNe~Ia.  In the specific models
studied in this paper, differences in the bolometric LCs are
relatively small, and attributable to the reduction in flux mean
opacity which accompanies the onset of Fe~II/Co~II lines
\citep{Pinto-Eastman_WLR,Kasen_NIR}.  The bolometric differences
become most prominent around 30 days after $B$-maximum, when the
brighter models show a slight rise to a secondary bolometric maximum.
Differences in diffusion time may therefore be somewhat more relevant
in formulations of the WLR employing later time data.

The bolometric LCs of SNe~Ia may also depend sensitively on the ejecta
kinetic energy and the radial distribution of \Nifs\ (see Paper~I).
By varying these two other parameters, it is possible to construct
sets of models that show greater or lesser degrees of scatter and/or
correlation in the bolometric WLR.  In addition, NLTE effects (not
included here) may also contribute to a stronger luminosity dependence
of the bolometric LCs, especially if the probability of wavelength
redistribution in lines is strongly temperature dependent.  Such
issues will be the subject of future studies.  However, because the
rate of color evolution dominates the $B$-band decline rate, the WLR
should remain a robust luminosity calibration even when the bolometric
LCs of SNe~Ia show significant random variations.

The bolometric rise time to maximum is the same for all models in this
paper: $\tbol = 15.5$~days. The \Bband\ rise time (\tb), on the other
hand, depends on the color evolution. Because iron group recombination
occurs earlier in lower \Mni\ models, their colors begin evolving
redward at earlier epochs.  Thus, in both models and observations, the
dimmer SNe~Ia exhibit a shorter \tb.

Because Fe~II/Co~II line blanketing generally increases to the blue,
the WLR will usually be stronger in bluer wavelength bands.  The
$V$-band, for instance, is impacted by the development of Fe~II/Co~II
lines to a lesser extent than the \Bband.  This leads, in both models
and observations, to a slower $V$-band decline rate \dmfv\ and a weaker
dependence of \dmfv\ on the peak $V$-band magnitude \Mv.

The \Bband\ decline rate also has an intimate connection to the
double-peaked morphology of the far red and NIR LCs of SNe~Ia.  The
flux absorbed in Fe~II/Co~II lines at blue wavelengths is
redistributed to the red, eventually leading to a NIR secondary
maximum
\citep{Pinto-Eastman_II, Kasen_NIR}.  
Naturally, in both models and observations, SNe~Ia with faster \Bband\
decline rates have earlier NIR secondary maxima.

On the whole, one recognizes that \emph{the ionization evolution of
iron group elements is the most significant physical factor in
understanding SN~Ia LCs and the WLR}.  The faster overall
spectroscopic and photometric evolution of dimmer SNe~Ia reflects,
primarily, their faster ionization evolution.

In addition, the distribution of iron group elements in the ejecta can
be identified as a crucial secondary parameter in the WLR.  All models
in this paper have substantial iron group abundance out to $v \approx
9000~\kms$. If the iron group elements are mixed to yet higher
velocities, the development of Fe~II/Co~II lines will be hastened and
intensified.  Thus, for given \Mni, enhancing the iron group abundance
in the outer layers of ejecta
\emph{increases} the \Bband\ decline rate.  The relatively low dispersion
in the observed WLR suggests that normal SNe~Ia exhibit limited
variability in their iron group distributions.  The physical origin of
this regularity, however, can only be addressed by detailed first
principle explosion modeling of SNe~Ia.

Fortunately for future precision cosmology experiments, the radiative
transfer effects identified in this paper can be used to monitor and
limit deviations in the WLR.  Variations in the velocity distribution
of iron group elements, for example, can be constrained by measuring
the Doppler-shifts of iron features in the SN~Ia spectra.  NIR
observations provide additional indicators of the ionization evolution
and iron group distribution in the ejecta \citep{Kasen_NIR}.  Close
inspection of well sampled color curves (or spectral time series)
should reveal equivalent or improved calibration techniques and may be
useful in quantifying subtle deviations from the standard
evolution. These and related tests should be feasible with the data
from ongoing observational campaigns, and will help to limit the
intrinsic scatter or evolutionary biases potentially affecting SNe~Ia
cosmology experiments.  With the aid of detailed transfer models such
as those presented here, refinements to the calibration techniques
need not rely on the fortuitous discovery of empirical correlations,
but can be devised to reflect the essential physics of SN~Ia light
curves.

\acknowledgements   
The authors would like to thank Phil Pinto and Ron Eastman for very
helpful discussions regarding the behavior of the bolometric light
curves, both in their prior theoretical studies and in the
observations.  We also thank Alex Conley, David Jeffery, and Adam
Riess for helpful comments and discussions.  DK is supported by the
Allan C. Davis fellowship at Johns Hopkins University and the Space
Telescope Science Institute.  SEW acknowledges support from NASA
(NNG05GG08G), and the DOE Program for Scientific Discovery through
Advanced Computing (SciDAC; DE-FC02-01ER41176) This research used
resources of the National Energy Research Scientific Computing Center,
which is supported by the Office of Science of the U.S. Department of
Energy under Contract No.  DE-AC03-76SF00098.

%\bibliography{../../BIBIN/MAIN.bib}

\clearpage

\end{document}